\documentstyle[prb,aps,epsf]{revtex}
\begin{document}

\draft

\twocolumn
[\hsize\textwidth\columnwidth\hsize\csname @twocolumnfalse\endcsname
\title{Nonlinear excitations in CsNiF$_3$ in
magnetic fields perpendicular to the easy plane}
\author{M.~Orend\'{a}\v{c},$^{1}$ A.~Orend\'{a}\v{c}ov\'{a},$^{1}$ 
E.~\v{C}i\v{z}m\'{a}r$,^{2}$ J.-H.~Park,$^{2}$ A.~Feher,$^{1}$ S.~J.~Gamble,$^{2}$ 
S.~Gab\'{a}ni,$^{3}$ K.~Flachbart,$^{3}$ J.~Karadamoglou,$^{4}$ 
M.~Poirier,$^{5}$ and M.~W.~Meisel$^{2}$}
\address{$^{1}$ Faculty of Science, P. J. \v{S}af\'{a}rik
University, Park Angelinum 9, 041 54 Ko\v{s}ice, Slovakia}
\address{$^{2}$ Department of Physics and Center for Condensed Matter
Sciences, University of Florida, Gainesville, FL  32611-8440, USA}
\address{$^{3}$ Institute of Experimental Physics, Slovak Academy of
Sciences, Watsonova 47, 043 53 Ko\v{s}ice, Slovakia}
\address{$^{4}$ Institut Romand de Recherche Num\'erique en Physique
des Mat\'eriaux (IRRMA), EPFL, 1015 Lausanne, Switzerland}
\address{$^{5}$ Centre de Recherche en Physique du Solide,
D\'{e}partement de Physique, Univerit\'{e} de Sherbrooke, J1K 2R1 Qu\'{e}bec, 
Canada}

\date{\today}
\maketitle

\begin{abstract}
Experimental and numerical studies of the magnetic field
dependence of the specific heat and magnetization of single crystals 
of CsNiF$_3$ have been performed at 
2.4~K, 2.9~K, and 4.2~K in magnetic fields up to 9~T 
oriented perpendicular to the easy plane.  The experimental results 
confirm the presence of the theoretically predicted double peak structure 
in the specific heat arising from the formation of nonlinear spin modes. 
The demagnetizing effects are found to be negligible, and the overall  
agreement between the data and numerical predictions is better than 
reported for the case when the magnetic field was oriented in the easy plane. 
Demagnetizing effects might play a role in generating  
the difference observed between theory and experiment in previous work 
analyzing the excess specific heat using the sine-Gordon model.
\end{abstract}

\pacs{75.40Cx, 75.40.Mg, 75.30.Et, 75.10.Pq}

\twocolumn
%\narrowtext
]

\section{Introduction}
Significant developments in solid state physics have occurred when 
nonlinear effects have been taken into consideration.  
Low dimensional magnets have proven to be excellent model systems 
for these studies since long-range ordering is suppressed, whereas 
short-range ordering, if sufficiently pronounced, gives rise to nonlinear behavior. 
In this respect, spin chains garner special importance, and the interest 
in these one-dimensional (1D) magnets was piqued by Mikeska's prediction 
of the contribution of thermally excited solitons to the dynamics of a planar 
spin chain.\cite{Mikeska}  These excitations are expected to appear in a 
symmetry breaking magnetic field oriented in the easy plane of the chain. 
Since the compound CsNiF$_3$ is a good realization of an $S = 1$ Heisenberg 
planar ferromagnetic chain, it has become one of the most intensively studied 
quasi-1D systems. Its magnetic properties in a magnetic field oriented 
in the easy plane have been investigated by different theoretical 
approaches \cite{Wesselinova,Papa} and various experimental techniques. 
More specifically, the presence of sine-Gordon solitons was assumed in the 
interpretation of the central peak in the energy spectrum of slow 
neutrons,\cite{Steiner} of the temperature and field dependences of the 
spin-lattice relaxation time studied by NMR,\cite{Goto} 
and of the excess magnetic specific heat.\cite{Ramirez}
In addition, the suppresion of the energy gap in the excitation
spectrum of CsNiF$_3$ in a magnetic field oriented in the easy plane
was analyzed using quantum sine-Gordon field theory. \cite{Reich}

Although the sine-Gordon model has been widely applied in the analysis of the data
obtained on several Heisenberg chain systems, the existence of sine-Gordon 
solitons has not been proven unambiguously. For example, the temperature and
magnetic field dependences of the intensity in the central peak observed 
by neutron scattering in CsNiF$_3$ is better described by the spin-wave 
prediction in comparison with a model assuming scattering from a 
dilute gas of solitons.\cite{Reiter}  Furthermore, 
reasonable quantitative agreement between the experimentally measured 
peak of the excess specific heat and the corresponding 
theoretical prediction could be obtained only after an artificial 
renormalization of the soliton energy,\cite{Ramirez} and this aspect has been 
the subject of considerable debate.\cite{Mouritsen,Pini}  
It is noteworthy that a similar situation was found for the $S = 5/2$ 
Heisenberg planar ferromagnet TMMC.\cite{Borsa}   
In other work, optical studies were initially interpreted in terms of 
magnon-soliton scattering,\cite{Cibert} but were subsequently shown to 
be explainable by multi-magnon process.\cite{Kumar}  
Recent theoretical progress in calculating the magnon dispersion and 
thermodynamic properties of CsNiF$_3$ using a transfer matrix 
renormalization group algorithm\cite{Papa} has allowed for the comparison 
of the experimental data with the exact theoretical predictions.  Nevertheless, 
using generally accepted values of the exchange coupling $J$, the single-ion 
anisotropy $D$, and the $g$ factor for CsNiF$_3$, only qualitative agreement 
was found for the magnetic field dependence of the specific heat.

Since solitons are predicted to appear when the magnetic field is oriented in 
the easy plane, less attention has been devoted to the situation when the 
field is oriented perpendicular to the easy plane.  Apart from the experimental study of 
the temperature dependence of the susceptibility\cite{Dupas} and the 
field dependence of the magnetization\cite{Delica} for such a field orientation, 
other work focused predominantly on mapping the magnetic 
phase diagram obtained from the anomalous behavior of the elastic 
constants.\cite{Poirier}  In addition, a theoretical treatment addressed the 
field induced instability of the three-dimensional (3D) antiferromagnetically 
ordered phase at $T < T_N = 2.7$~K.\cite{Pich}

In fact, the situation is significantly richer than initially expected due to 
the formation of nonlinear spin modes when the magnetic field is orientated 
perpendicularly to the easy plane.\cite{Papa} More specifically, nonlinear effects 
lead to the formation of a characteristic double peak in the magnetic field 
dependence of the specific heat in the vicinity of a critical field $B_c$ 
at which the ground state is changed.  The existence of the double peak has 
already been predicted in $S = 1/2$ XY chains as well as $S = 1/2$ Heisenberg 
chains with Ising anisotropy.\cite{Papa1,Papa1a}  The proposed behavior of the 
specific heat significantly differs from that obtained in the dilute-magnon 
approximation, where a single broad maximum at $B_c$ is predicted.  
The motivation of our experimental study of CsNiF$_3$ was to search for the 
existence of the predicted double peak structure in the field dependence 
of the specific heat and to examine the thermodynamic response at several  
temperatures near $T_N$.
As we will present in the following sections, our results confirm the existence 
of the theoretically predicted double peak structure.  Moreover, the 
numerical analysis of our results indicates that the previously reported 
discrepancies between the excess specific heat data and the corresponding 
theoretical predictions may be partially attributed to demagnetization effects.  

\section{Experimental Details}
The single crystals of CsNiF$_3$ used for our work were light green 
and transparent throughout their volume, albeit with some internal 
cleavage planes noticeable. Due to the highly hydroscopic nature of the material, 
all samples were kept in paraffin oil until they were ready to be mounted.
For the specific heat study, a single crystal of mass 8.44~mg and  
approximate dimensions $7 \times 0.5 \times 0.5$~mm$^3$ was used.  This specimen 
was attached to a simple homemade microcalorimeter using Epotek E4110 
silver epoxy. The calorimeter consisted of a sapphire disk that supported 
a heater and a thermometer, both made from RuO$_2$ chip resistors.  
 The thermometer was calibrated against a commercial 
Cernox thermometer (Lakeshore CX-1030CD) which was located near the 
calorimeter.  The position of the calorimeter was fixed by a holder made 
of Vespel SP-1 that was needed to prevent motion in the magnetic field. 
The holder and the manganin wires from the thermometer and heater served 
as the thermal link to the stable reservoir. 
At 4.2~K, the thermal conductivity of the manganin wires (0.15~$\mu$W/K) 
represented only about 5\% of that of the vespel holder (3~$\mu$W/K).   
Consequently, the resultant value of the total thermal conductivity can be 
considered to be field independent.  The resistance of the thermometer was 
measured using a Wheastone bridge, with a PAR~124A lock-in amplifier serving 
as a null detector. While monitoring the cooling curve of a measurement, 
the analog output of the lock-in was read by a HP 3457A multimeter. 
The same device was used for measuring the voltage on the heater during the 
heating step, and the current was provided by a Keithley Model~220 
current source. Consequently, the temperature and magnetic
field dependences of the heater resistance did not influence the accuracy 
of the measurements. In the measuring cycle, the relative change of the
temperature of the sample ranged from 3\% to 5\%.
 The specific heat was calculated from the ratio of the 
characteristic cooling relaxation time and the thermal conductivity of the link.  
We estimate the overall accuracy of the measurement to be better than 5\%.
  
For the estimation of the correction due to demagnetizing effects, the field 
dependence of the magnetization was studied up to 7~T at several selected 
temperatures using a Quantum Design \textsc{SQUID} magnetometer.  
For these measurements, a single crystal, with dimensions similar to those 
used for specific heat studies, was placed in a polyethylene vial 
that was held by a straw.  The background contribution from the vial and straw 
was independently measured to be less than 1\% of the total signal and 
was subtracted.

\section{Results}
For our first step, the specific heat of CsNiF$_3$ was studied from 2~K
to 5~K in zero magnetic field. These results were compared with 
the data reported by other workers,\cite{Lebesque} and the comparison is 
presented in Fig.~1. Both sets of data are characterized by a 
$\lambda$-like anomaly at about 2.7~K, indicating the presence of long-range 
ordering. However, the larger peak observed in our experiment 
suggests that our single crystal specimen was of higher quality than the one 
used in the other work.\cite{Lebesque}  The excellent agreement of the 
critical temperatures confirms that the silver epoxy did not
deteriorate the bulk properties of our sample. The difference between 
both data sets can be described by the equation
\begin{equation}
C(T)\;=\;aT\;+\;bT^3\;\;\;\;,
\end{equation}
when $a=0.9 \pm 0.3$ J/K$^2$ and $b=0.88 \pm 0.03$ J/K$^4$. This difference
is attributed to the specific heat of the addenda created predominantly by
the vespel holder and the sapphire substrate.
In addition, since the possible amount of paramagnetic impurities in the silver 
epoxy is at most 0.001\%,\cite{epoxy} the specific heat of the addenda should 
have a negligible magnetic field dependence.  
The fact that the lattice contribution of CsNiF$_3$ was not subtracted
from the total heat capacity does not compromise the
evaluation of the magnetic field dependence of the excess specific heat, the
quantity of primary interest in this study.

The magnetic field dependence of the specific heat was studied at 2.4~K, 2.9~K, 
and 4.2~K, in fields up to 9~T. As mentioned previously, the weak 
magnetoresistance of the heater is simple to accommodate and has no 
direct influence in a thermal relaxation study.  Similarly, the weak 
magnetoresistance of the thermometer is negligible since it does not influence  
the thermal relaxation time of the calorimeter.
On the other hand, the data were corrected using the known magnetoresistance 
of the Cernox thermometer.\cite{Cernox}  As expected, the correction was most 
significant at $T=2.4$~K in magnetic fields higher than 6~T, where it represented 
3.5\% of the measured heat capacity value. The resultant magnetic field dependence 
of the total specific heat is presented in Fig.~2. The $\lambda$-like anomaly observed
at 2.4~K in a field of about 2~T corresponds to field-induced long-range ordering. 
The values of the critical temperature and field agree well with the 
magnetic phase diagram obtained from ultrasonic investigations of specimens 
from the same batch of crystals.\cite{Poirier}

\section{Discussion}
For the investigation of the effect of the nonlinear 
excitations on the equilibrium thermodynamic properties, most of the attention 
has concentrated on the excess isothermal specific heat
\begin{equation}
\Delta C(T,B)\;=\;C(T,B)\;-\;C(T,0)\;\;\;,
\label{excess}
\end{equation}
studied as a function of magnetic field $B$ at a temperature $T$.  
The numerical predictions\cite{Papa} of the excess specific heat 
for CsNiF$_3$ in a magnetic field oriented perpendicular to the easy plane were 
derived from a 1D model described by the $S$ = 1 Hamiltonian
\begin{equation}
{\cal H} \,=\, -J \sum_{i} {S_{i} \cdot  {S_{i+1}} \,+\,D \sum_{i
}{({S_i}^z)}^{2} \,+\,g\,\mu_{B}\,B\sum_{i} {S_i}^z}\, ,
\label{ham}
\end{equation}
using the values $J/k_B = 23.6$ K, $D/k_B = 8.25$ K, and $g = 2.13$. The 
numerical simulations were performed for temperatures 2.4~K and 4.2~K,  
the former being below the critical temperature of the real material. 
The results of the numerical simulations suggest that the predicted 
double peak structure in the excess specific heat becomes more 
pronounced at lower temperatures. Consequently, in the corresponding 
experimental study, the influence of interchain 
correlations should be considered in the critical region.

The ratio of intrachain and interchain exchange coupling constants 
($J/J^{\prime} \approx 500$) makes CsNiF$_3$ one of the best representatives 
of a quantum spin chain.\cite{Baehr}  Indeed, the long-range ordering 
observed at $2.66 \pm 0.01$~K in this study is manifested as a small spike located on the
broad maximum appearing due to the short-range correlations. In addition, 
the amount of entropy removed by the spike itself ($\approx 0.05$ J/(K mol)) 
is by far less than 1\% of the total amount of entropy for the $S = 1$ system, 
namely 9.13~J/(K mol). In such a situation, it appears reasonable 
to approximate the contribution of the interchain coupling to the total specific 
heat by fitting the temperature dependence of the specific heat using data below 
and above the spike, \emph{i.e.} outside the critical region.  
The result of this approximation is shown as the solid line in Fig.~1, and 
the enhancement of the specific heat 
in the critical region due to the interchain correlations is apparent
(see inset in Fig. 1).
Such an enhancement generates a deviation from the corresponding 
theoretical predictions,\cite{Papa} which were made for a purely 1D system. 
Consequently, when comparing the experimental excess specific heat data 
obtained at 2.4~K with the theory, the specific heat value calculated from the 
aforementioned fit at 2.4~K was taken as the reference value $C(T,0)$ in 
Eq.~(\ref{excess}).  As a result, if the experimental specific heat data are 
represented by $C(T,B)$ in Eq.~(\ref{excess}), then the quantity $\Delta C$ should 
tend to a nonzero value in the zero field limit.  In other words, in the critical 
region, the interchain coupling is responsible for the finite value of the excess 
specific heat. The comparisons of the experimental excess specific heat data with 
the numerical predictions are presented in Fig.~3. As expected at 2.4~K, 
the agreement between the theory and experiment is greatly 
improved in magnetic fields greater than $\approx 3$~T, which is 
high enough to overcome the influence of the long-range correlations 
and decompose the system into independent chains.  
This difference between the experimental data and theoretical predictions 
persists, but is smaller, at 2.9~K, and it disappears at 4.2~K. 
It should be noted that critical magnetic field observed at 2.4~K 
is consistent with the phase diagram obtained from 
the ultrasonic study which used specimens from the same batch of 
crystals.\cite{Poirier}  
Furthermore, the experimental data reproduce the predicted positions of the 
minimum values of $\Delta C$, as well as the shift of the position
of the second maximum towards lower fields with decreasing temperature.  
Although some quantitative differences persist, it should be stressed
that the overall agreement between the numerical predictions and
experimental data is better than that achieved when the magnetic field is 
oriented in the easy plane.\cite{Papa}

Several potential reasons might explain the observed differences. 
For example, quantitative comparisons between the experimental data and the 
theoretical predictions should take demagnetizing effects into account.  
The correction of the specific heat due to demagnetizing effects 
can be calculated as follows.\cite{Levy}  The internal magnetic field is
written as
\begin{equation}
B_i\;=\;B_{e}\;-\;NM\;\;\;,
\end{equation}
where $B_i$ and $B_{e}$ stand for the internal and external fields, 
while $M$ and $N$ denote the magnetization and demagnetizing 
factor.  The specific heat of the sample may then be written as
\begin{equation}
C_{B_i}(T,B_i)=C_{B_{e}}(T,B_i)+\frac{[NT({\partial M}/{\partial
T})^2_{B_{e}}]}{[1-N({\partial M}/{\partial B_{e}})_T]}\;\;\;.
\end{equation}
The needle shape of the sample, which was oriented with its long axis 
parallel with the applied external magnetic field, 
should lead to a small value of the demagnetizing factor. 
Indeed, for the dimensions of our single crystal specimen, the average 
demagnetizing factor is approximately 0.04 in SI units. In order to evaluate 
$(\partial M/ \partial B_e)_T$, the field dependence of magnetization was 
investigated up to 7~T at 2.4~K, 2.9~K, and 4.2~K. For clarity, only the 
results obtained at 2.4~K and 4.2~K are presented in Fig.~4.  For all 
temperatures, the magnetization is characterized by an 
increase up to the critical field, followed by the tendency to 
saturate at higher fields. It should also be noted that, for the 
magnetization data corrected for demagnetizing effects, 
some deviation from the numerical predictions appears.
The values of the term $(\partial M/ \partial T)_{B_{e}}$ 
were evaluated with the help of the additional field dependences 
of the magnetization studied at temperatures about 2\% higher and lower 
than the aforementioned ones. Then $(\partial M/ \partial T)_{B_{e}}$ for 2.4~K, 
2.9~K, and 4.2~K were approximated by the corresponding differences of the 
magnetization at the neighboring temperatures. The results of this analysis are 
presented in Fig.~5. The remarkable feature of the 
$(\partial M/ \partial T)_{B_{e}}$ quantity is its nontrivial field 
dependence in the region of low fields, where it is most pronounced at 2.4~K
and also observed at 2.9~K, but is completely absent at 4.2~K. Taking into 
account that both 2.4~K and 2.9~K belong to the critical region of 
temperatures for CsNiF$_3$, the observed behavior can be attributed to the 
effect of interchain coupling. It should be noted that, for 2.4~K and 2.9~K, 
the value of the field, where the anomalous behavior of 
$(\partial M/ \partial T)_{B_{e}}$ 
disappears ($\approx 3$ T), agrees very well with that for 
which $\Delta C$ starts to follow the theoretical prediction for a pure 1D 
system. Such agreement supports the suggestion that $\approx 3$~T is sufficient to 
suppress the 3D effects near $T_N$.   
After evaluating $(\partial M/\partial B)_T$ and 
$(\partial M/ \partial T)_{B_{e}}$, the correction to the specific heat 
due to the demagnetizing effects was calculated. 
However, it turns out that the correction does not exceed 0.1\% of the 
uncorrected specific heat value, thus demagnetizing effects can not be 
responsible for the observed differences. This fact contrasts with  
the situation when magnetic field is oriented in the easy plane, where 
such a correction was proven to be significant.\cite{Ramirez1}

Naturally, tuning the parameters $J$, $D$, and $g$ might improve 
the agreement between the numerical predictions and the experimental results.  
Consequently, the susceptibility,\cite{Dupas} magnetization,\cite{Delica} 
and present specific heat data were reanalyzed using the transfer matrix 
renormalization group technique. Since no evidence for
the change of exchange coupling constant was found in the previous
analysis,\cite{Papa} the value $J/k_B = 23.6$~K was adopted in the 
recalculation. In the new set of parameters, the `standard' value of 
$D/k_B =9$~K was chosen, whereas $g=2.19$ was obtained from fitting 
the susceptibility\cite{Dupas} and magnetization\cite{Delica} with the 
new $D$ value.  It should be noted that such a choice of 
new $D$ and $g$ parameters does not significantly change 
the magnitude of the critical field, $B_c=D/g \mu_B \approx 6$~T 
as suggested by the specific heat data. However, as can be seen in 
Fig.~3, the new predictions are shifted towards larger fields and 
the values of $\Delta C$ are increased, thereby making the agreement with the 
experimental data worse than for the original set of parameters. 
Further increasing of the $D$ value leads to unsatisfactory 
fitting of the susceptibility and magnetization. Consequently,  
tuning the parameters in the framework of a pure 1D model 
does not suppress the differences between the numerical prediction and the 
experimentally measured excess specific heat.

Finally, the possible effect of the tilting of the c-axis with respect to the 
direction of the magnetic field may be considered.  It is not 
straightforward to quantitatively evaluate this effect, but   
it can be expected to influence the experimental results. For example,
the susceptibility below 8 K
is about five times higher for magnetic field oriented in the easy plane 
than that for the perpendicular orientation.\cite{Papa}
Although the magnitudes of $\Delta C$ are comparable for the different 
field orientations, the corresponding field dependences are completely 
different. Thus, tilting may nontrivially contribute to the differences 
between the numerical predictions and experimental specific heat data.

\section{Conclusions}
The experimental study of the magnetic field dependence of excess specific heat 
performed at 2.4~K, 2.9~K, and 4.2~K has confirmed the theoretically predicted 
double peak structure reflecting the nonlinear behavior of the system. 
The excess specific heat data reproduce the predicted 
shift of the second maximum towards lower fields with decreasing temperature. 
It was found that, similarly as for the orientation of the magnetic 
field in the easy plane, it is not possible to fit all available 
thermodynamic data using a pure 1D model with a single set of parameters. 
Although tilting of the sample may be an alternative explanation of the
persisting deviations, demagnetizing effects are determined to be negligible.
This conclusion contrasts with the one made in  
Ref.~\onlinecite{Ramirez}, where the magnetic field was oriented in the easy 
plane and where the correction for the shape dependence had a pronounced effect. 
However, even with the incorporation of the demagnetization correction, 
the agreement between the excess specific heat data\cite{Ramirez} and the 
corresponding exact numerical predictions\cite{Papa} is worse than in our case. 
In Ref.~\onlinecite{Ramirez}, the estimation of 
$(\partial M/ \partial T)_{B_{e}}$ from the field dependence of the 
magnetization\cite{Rosinski} might not be sufficiently accurate due to the 
restricted amount of data that is available.  Indeed, in situations where the 
demagnetizing effects are pronounced, more detailed mappings of the field and 
temperature dependences of the magnetization were performed.\cite{Landau} 
Naturally, a question may arise about the influence of demagnetizing effects 
in the specific heat studies of other soliton - bearing systems. 
A considerable amount of theoretical effort\cite{Tinus,Tinusa,Tinusb} 
has been developed to understand the differences and to improve the agreement 
with the specific heat data.\cite{Borsa,Kopinga1}  However, to the best of our 
knowledge, the data themselves were not corrected for demagnetizing  effects.
Consequently, a future experiment might focus on clarifying the role of the 
demagnetizing effects in CsNiF$_3$ as a soliton - bearing system.    
Furthermore, since the predicted double peak structure seems to be a generic 
feature of the nonlinear behavior in classical and quantum spin chains, 
future experimental effort may focus on measuring $\Delta C$ in other 
related materials.

\acknowledgments
We would like to thank N.~Papanicolaou for enlightening discussions and for 
providing us with the results of numerical calculations published in
Ref. \onlinecite{Papa}.  Early contributions
from B.~C.~Watson during the design and construction of the probe, and 
B. Andraka during the initial phases of the work, are gratefully 
acknowledged.  This work was supported, in part, by the National Science 
Foundation (INT-0089140, DMR-0113714, DGE-0209410, and DMR-0305371) and the 
Slovak Grant Agency (VEGA 1/0430/03).

\begin{figure}
\epsfverbosetrue
\vspace{10mm}
\epsfxsize=8cm
\epsffile{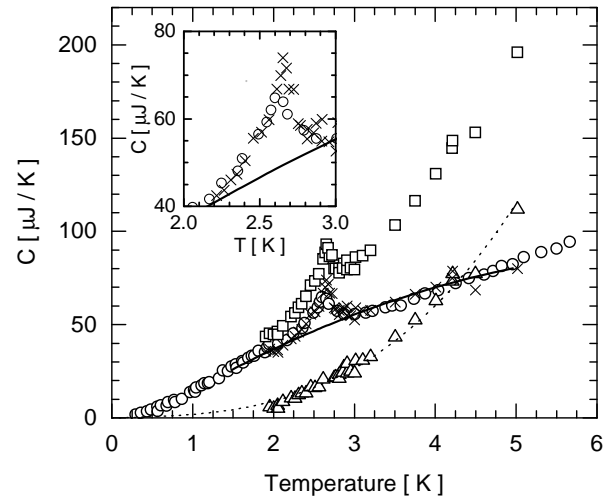}
\vspace{2mm}
\caption{The temperature dependence of total heat capacity of CsNiF$_3$ together 
with the addenda (squares) is compared to the data of magnetic heat 
capacity reported in Ref.~20 (circles), after renormalization to the mass 
of our sample.  The estimated contribution of the addenda for the present study
is shown by the triangles.  The heat
capacity of our sample, i.e. the difference between the squares and triangles, 
is represented by the crosses.  A least square fit 
of the lattice contribution and the addenda is denoted by the dotted line, see 
Eq.~(1). The solid line is obtained by fitting the data outside the critical region, 
see the text for a more detailed discussion.  The inset provides an expanded view 
near the critical region.}
\label{fig. 1}
\end{figure}

\begin{figure}
\epsfverbosetrue
\vspace{10mm}
\epsfxsize=8cm
\epsffile{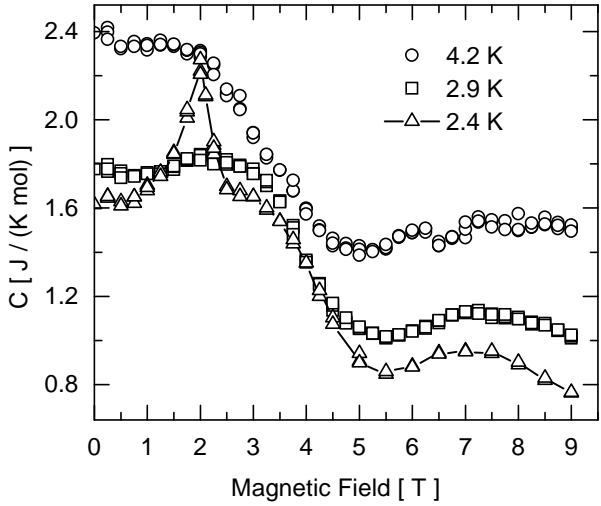}
\vspace{2mm}
\caption{The magnetic field dependence of the total specific heat of CsNiF$_3$ is shown
at 2.4~K (triangles with lines as guides for the eyes), 
2.9~K (squares) and 4.2~K (circles).}
\label{fig. 2}
\end{figure}

\begin{figure}
\epsfverbosetrue
\vspace{10mm}
\epsfxsize=9cm
\epsfysize=18cm
\epsffile{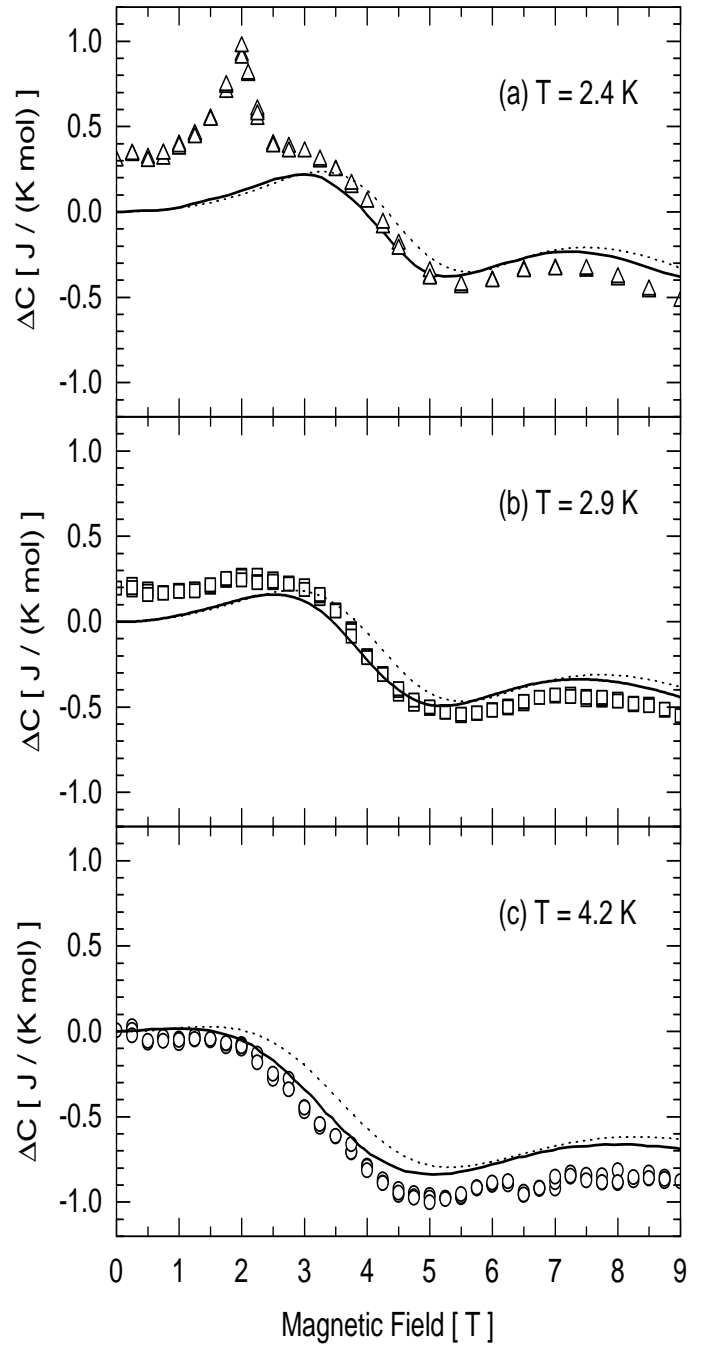}
\vspace{2mm}
\caption{The magnetic field dependence of $\Delta C$, Eq. (2), at
(a) $T = 2.4$~K, (b) $T = 2.9$~K, and (c) $T = 4.2$~K, are compared with 
theoretical expectations.  The solid lines represent the numerical predictions 
when $J/k_B = 23.6$~K, $D/k_B = 8.25$~K, and $g = 2.13$, while the dotted lines are for 
$J/k_B = 23.6$~K, $D/k_B =9$~K, and $g = 2.19$.}
\label{fig. 3}
\end{figure}

\begin{figure}
\epsfverbosetrue
\vspace{10mm}
\epsfxsize=8cm
\epsffile{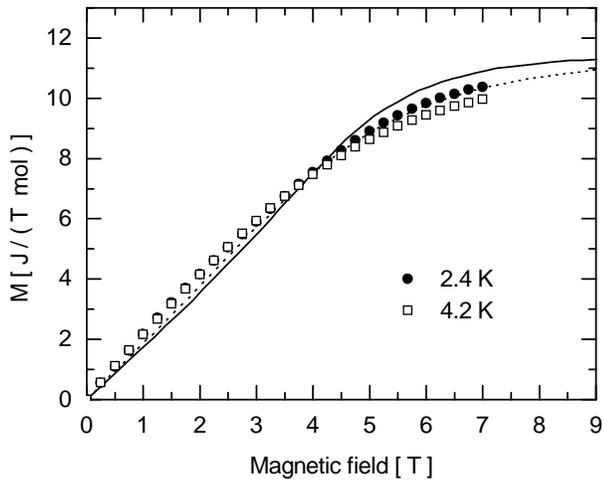}
\vspace{2mm}
\caption{Magnetic field dependence of magnetization of CsNiF$_3$ studied
at 2.4~K (squares) and 4.2~K (circles). The solid and dotted 
lines represent the numerical predictions at 2.4~K and 4.2~K, respectively, 
when $J/k_B = 23.6$~K, $D/k_B = 8.25$~K, and $g=2.13$.}
\label{fig. 4}
\end{figure}

\begin{figure}
\epsfverbosetrue
\vspace{10mm}
\epsfxsize=8cm
\epsffile{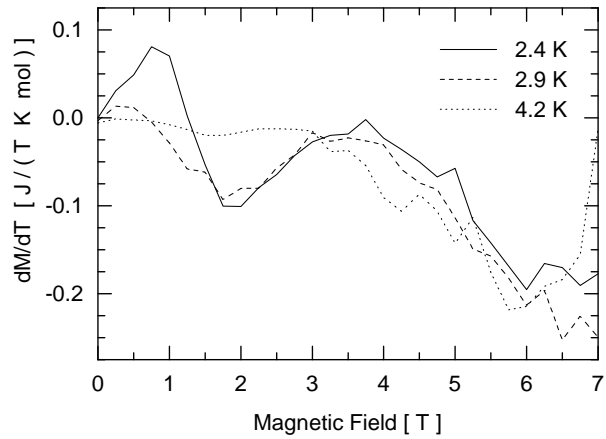}
\vspace{2mm}
\caption{Magnetic field dependence of $(\partial M/\partial T)_{B_{e}}$
calculated at 2.4~K (solid line), 2.9~K (dashed line), and 4.2~K (dotted line).}
\label{fig. 5}
\end{figure}

\end{document}